\documentclass[conference,a4paper]{IEEEtran}
\IEEEoverridecommandlockouts
\usepackage{cite}
\usepackage{amsmath,amssymb,amsfonts}
\usepackage{algorithmic}
\usepackage{graphicx}
\usepackage{textcomp}
\usepackage{xcolor}
\usepackage{multirow}
\usepackage{multicol}
\usepackage{subcaption}
\usepackage{textcomp}
\usepackage[absolute]{textpos}
\def\BibTeX{{\rm B\kern-.05em{\sc i\kern-.025em b}\kern-.08em
    T\kern-.1667em\lower.7ex\hbox{E}\kern-.125emX}}
\newcommand{\copyrightstatement}{
    \begin{textblock}{15}(0.3,0.2)    
         \noindent
         \centering
         \textblockcolour{white}
         \footnotesize
         \copyright 2024 IEEE. Personal use of this material is permitted. Permission from IEEE must be obtained for all other uses, in any current or future media, including reprinting/republishing this material for advertising or promotional purposes, creating new collective works, for resale or redistribution to servers or lists, or reuse of any copyrighted component of this work in other works.
    \end{textblock}
}
\begin{document}

\copyrightstatement
\title{Towards Video Codec Performance Evaluation: A Rate-Energy-Distortion Perspective\\
}
\author{\IEEEauthorblockN{Geetha Ramasubbu, \ Andr\'e Kaup, \ Christian Herglotz }
\IEEEauthorblockA{\textit{Multimedia Communications and Signal Processing,} \\
\textit{Friedrich-Alexander University Erlangen-Nürnberg (FAU),}\\
\textit{Erlangen, Germany} \\
{\{geetha.ramasubbu, andre.kaup, christian.herglotz\}@fau.de.}}
}
\maketitle

\begin{abstract}
The Bjøntegaard Delta rate (BD-rate) objectively assesses the coding efficiency of video codecs using the rate-distortion (R-D) performance but overlooks encoding energy, which is crucial in practical applications, especially for those on handheld devices. Although R-D analysis can be extended to incorporate encoding energy as energy-distortion (E-D), it fails to integrate all three parameters seamlessly. This work proposes a novel approach to address this limitation by introducing a 3D representation of rate, encoding energy, and distortion through surface fitting. In addition, we evaluate various surface fitting techniques based on their accuracy and investigate the proposed 3D representation and its projections. The overlapping areas in projections help in encoder selection and recommend avoiding the slow presets of the older encoders (x264, x265), as the recent encoders (x265, VVenC) offer higher quality for the same bitrate-energy performance and provide a lower rate for the same energy-distortion performance.

\end{abstract}

\section{Introduction}
Traditional video coding aims to encode content at the lowest bit rate with minimal distortion. Distortion is assessed through objective (e.g., peak signal-to-noise ratio, PSNR \cite{HuynhThu2008}) and subjective evaluations. Bitrate is the term that has been used to describe transmission or storage requirements since the standardization of the first video codec \cite{CCITT1988}. Despite the goal of video coding being to obtain the lowest possible bitrate and minimal distortion, it is impossible to minimize both rate and distortion simultaneously. Therefore, a trade-off between bitrate and distortion is characterized by rate-distortion (R-D) curve fitting.

Gisle Bjøntegaard introduced the Bjøntegaard-Delta (BD) metric in 2001 \cite{Bjoentegaard01}, which metric utilizes a third-order logarithmic polynomial to model the bitrate and PSNR performance, approximating a given rate-distortion (R-D) curve. It effectively quantifies the average difference between two R-D curves, and the BD metric offers a robust evaluation of coding efficiency. Consequently, it is a widely used evaluation metric in video coding research. However, the BD metric does not account for the energy demand of the encoders. Notably, modern codecs provide many compression methods, increasing the processing complexity of the encoders \cite{VVCComplexity}, resulting in a considerable increase in the energy demand. This substantial energy increase presents a significant challenge for practical video applications \cite{Carroll13}. Though the BD metric has been instrumental in evaluating coding efficiency, there's a pressing need to consider energy efficiency \cite{Herglotz2022_SweetStreams} alongside traditional performance metrics. With the rising importance of energy-aware video coding, there is a need for comprehensive evaluation metrics that encompass both coding and energy efficiency aspects. 

Although we can extend R-D analysis to include encoding energy as an energy-distortion (E-D) analysis \cite{Herglotz2020d}, where conceptually, the bitrate is replaced by energy. However, this approach still falls short of fully incorporating all three parameters harmoniously. To address this challenge, we propose the utilization of a 3D representation that encapsulates rate, encoding energy, and distortion through surface fitting techniques. By employing this multidimensional representation, we aim to provide a comprehensive evaluation framework that enables a more holistic understanding of video coding performance.

In our paper, we first measure the encoding energy for various presets and three different encoder implementations. Then, we explore various surface fitting techniques to assess their effectiveness in accurately capturing the relationships between rate, encoding energy, and distortion. Additionally, we offer interpretations derived from the proposed 3D representation, shedding light on the relation between these parameters and their implications for video coding efficiency. With this approach, we aim to evaluate based on a perspective that is both comprehensive and nuanced, going beyond the confines of traditional R-D analysis. Our findings help to identify efficient encoding configurations based on rate, encoding energy, and distortion. 

The rest of the paper is structured as follows:  Section \ref{sec:setup} introduces the encoding configurations and sequences used in the study, along with the encoding energy measurement setup, followed by a 3D R-E-D representation in Section \ref{sec:fitting}, where we extend the rate by encoding energy here as $D$ is the function to be minimized. Also, we introduce the surface fitting techniques used for 3D R-E-D representation in Section \ref{sec:fitting}. Then, in Section \ref{sec:eval}, we evaluate the surface fitting techniques and present various interpretations based on 2D projections of the 3D representation in Section \ref{sec:discussion}.


\section{Experimental Setup} \label{sec:setup}
Our work uses three different software encoder implementations x264 \cite{x264}, x265 \cite{x265}, and VVenC \cite{VVenC},  on an Intel Xeon processor to perform multi-core encoding, where we encode the first 64 frames of each sequence at various presets $\textit{ultrafast}$, $\textit{veryfast}$, $\textit{fast}$, $\textit{slow}$, $\textit{veryslow}$ and various constant rate factor (CRF) values, 18, 23, 28, 33 for x264 and x265 encoder implementations. For VVenC, we use $\textit{faster}$, $\textit{fast}$, $\textit{medium}$, $\textit{slow}$, $\textit{slower}$ presets \cite{VVenC}, and various quantization parameter (QP) values, 22, 27, 32, 37. As in \cite{Herglotz18c}, we describe the energy demand of the encoding process as a difference between the total energy consumed $E_{\mathrm {total}}$ during the encoding process and the energy consumed in idle mode $E_{\mathrm {idle}}$ over the same encoding duration $T$ as in \eqref{ee}.
\begin{equation}\label{ee} E_{\mathrm {enc}} = E_{\mathrm {total}}- E_{\mathrm {idle}} \end{equation}
In this work, we used a desktop PC with an Intel Xeon CPU with 16 cores and CentOS 8 as an operating system (OS). We employed the integrated power meter in Intel CPUs, running average power limit (RAPL) \cite{RAPLinAction2018}, that directly returns aggregated energy values $E_{\mathrm {total}}$ and $E_{\mathrm {idle}}$. Furthermore, we repeated measurements and 
we performed the confidence interval test proposed in \cite{Bendat1971}, to ensure the statistical significance of the measured encoding energies \cite{Ramasubbu22}. We consider 22 sequences from the JVET common test conditions \cite{CTC} with various sequence characteristics such as frame rate, resolution, and content. Ultimately, we generated 20 supporting points (4 CRFs x 5 Presets) per sequence for all three encoders, x264, x265, and VVenC. These supporting points, triplets ($r_{\mathrm {s}}, e_{\mathrm {s}}, d_{\mathrm {s}}$) consisting of rate, energy, and distortion values, are then used in Section \ref{sec:fitting} to construct R-E-D surfaces.
\section{Rate-Energy-Distortion surface fitting} \label{sec:fitting}
Building upon the concept of the rate-distortion (R-D) function, the rate-energy-distortion function $D(R, E)$ can be defined as the achievable distortion D for a given rate-energy pair. To do surface fitting, we use various interpolation techniques. In our case, interpolation involves creating a distortion function $D$ that matches given data values $d_i$ at given data points,  $r_i = \log{R_i}$ and  $e_i = \log{E_i}$, which are respective logarithmic rate and energy values, where $D(r_i, e_i) = d_i$, for all i, where i  represents a tuple of CRF/QP and preset. In this work, first, we used piece-wise linear interpolation to perform the surface fitting. Secondly, we employ a polynomial function similar to \cite{Li10}, that describes $D$ as a function of logarithmic rate and energy values, $r_i$ and $e_i$, with model coefficients $\{p_0,...,p_5\}$ as follows: 
\begin{equation}\label{custom}
\small
\begin{split}
    D(r,e) = p_0 + p_1 \cdot r^3 + p_2 \cdot r^2 + p_3 \cdot r + p_4\cdot e^2 +p_5\cdot e
\end{split}
\vspace{-0.5cm}
\end{equation}
Lastly, we used mixed polynomial model, where we define the distortion $D$ as a function of logarithmic rate and energy values, $r_i$ and $e_i$, with model coefficients $\{p_0,...,p_8\}$ as follows:
\begin{equation}\label{poly32}
\small
\begin{split}
    D(r,e) = p_0 + p_1 \cdot r + p_2 \cdot e + p_3 \cdot r^2 + p_4 \cdot r \cdot e + \\
    + p_5\cdot e^2 +p_6\cdot r^3 + p_7 \cdot r^2\cdot e + p_8\cdot r\cdot e^2
\end{split}
\end{equation}

\begin{table}[]
\centering
\caption{Evaluation metric MAPE for supporting points and non-supporting points.}
\label{tab:accuracy}
\begin{tabular}{|l|llllll|}
\hline
\multirow{3}{*}{\textbf{Enc Imp}} & \multicolumn{6}{c|}{\begin{tabular}[c]{@{}c@{}}\textbf{Mean absolute percentage error,}\\ (MAPE \%)\end{tabular}}                                      \\ \cline{2-7} 
                         & \multicolumn{3}{l|}{\textbf{supporting points, ${\epsilon_\mathrm{s}}$}}                                         & \multicolumn{3}{l|}{\textbf{non-supporting points, ${\epsilon_\mathrm{ns}}$}}                   \\ \cline{2-7} 
 &
  \multicolumn{1}{l|}{\textbf{Linear}} &
  \multicolumn{1}{l|}{\begin{tabular}[c]{@{}l@{}}\textbf{Poly}. \\ \eqref{poly32}\end{tabular}} &
  \multicolumn{1}{l|}{\begin{tabular}[c]{@{}l@{}}\textbf{Poly}.\\ \eqref{custom}\end{tabular}} &
  \multicolumn{1}{l|}{\textbf{Linear}} &
  \multicolumn{1}{l|}{\begin{tabular}[c]{@{}l@{}}\textbf{Poly}. \\ \eqref{poly32}\end{tabular}} &
  \begin{tabular}[c]{@{}l@{}}\textbf{Poly}.\\ \eqref{custom}\end{tabular} \\ \hline
x264                     & \multicolumn{1}{l|}{0} & \multicolumn{1}{l|}{0.62} & \multicolumn{1}{l|}{1.03} & \multicolumn{1}{l|}{0.56} & \multicolumn{1}{l|}{0.94} & 1.15 \\ \hline
x265                     & \multicolumn{1}{l|}{0} & \multicolumn{1}{l|}{0.40} & \multicolumn{1}{l|}{0.64} & \multicolumn{1}{l|}{0.48} & \multicolumn{1}{l|}{0.64} & 0.72 \\ \hline
VVenC                    & \multicolumn{1}{l|}{0} & \multicolumn{1}{l|}{0.11} & \multicolumn{1}{l|}{0.19} & \multicolumn{1}{l|}{-}    & \multicolumn{1}{l|}{-}    & -    \\ \hline
\end{tabular}
\end{table}
\begin{figure*}
\begin{multicols}{3}
    \includegraphics[width=\linewidth]{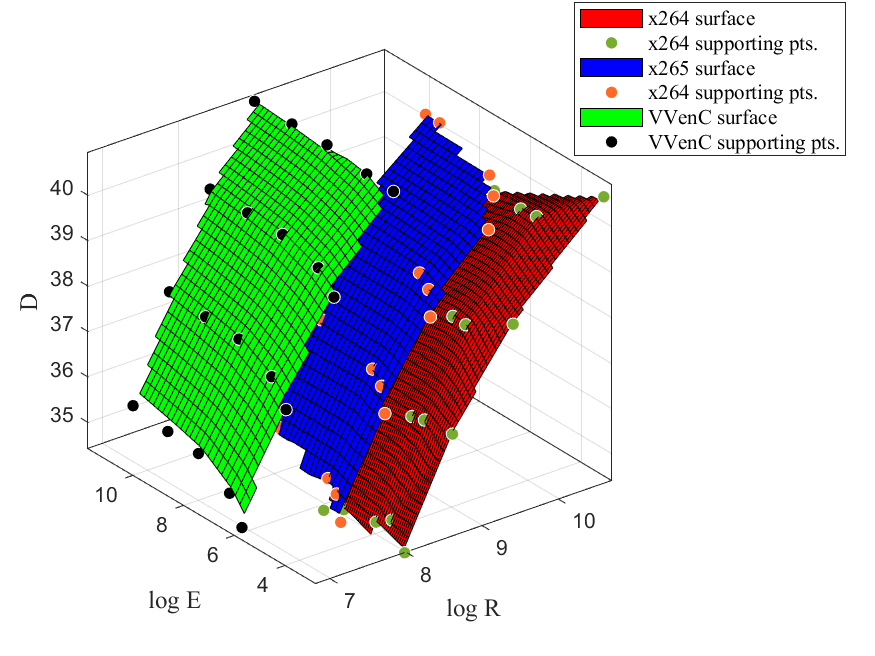}\par 
    \includegraphics[width=\linewidth]{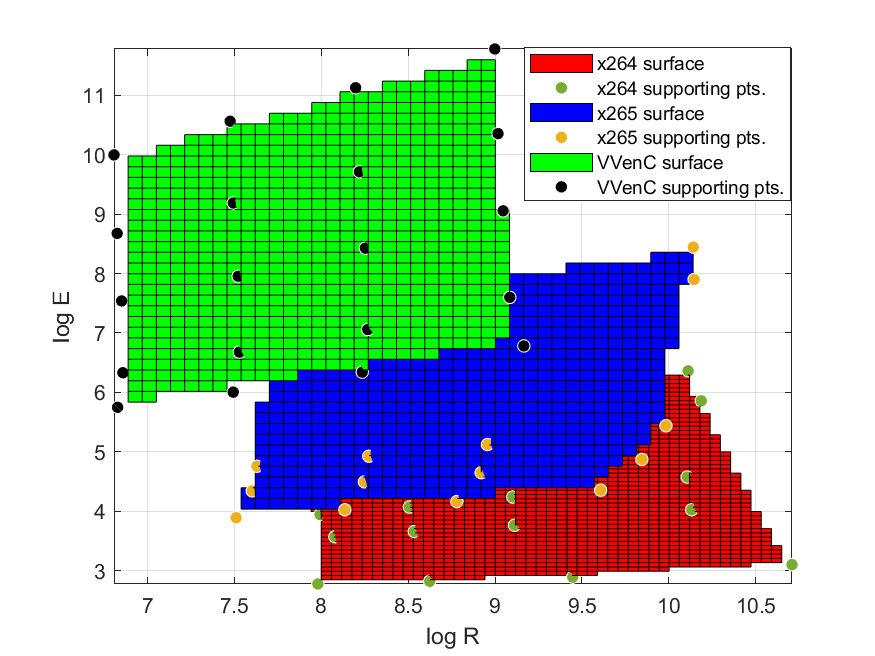}\par 
    \includegraphics[width=\linewidth]{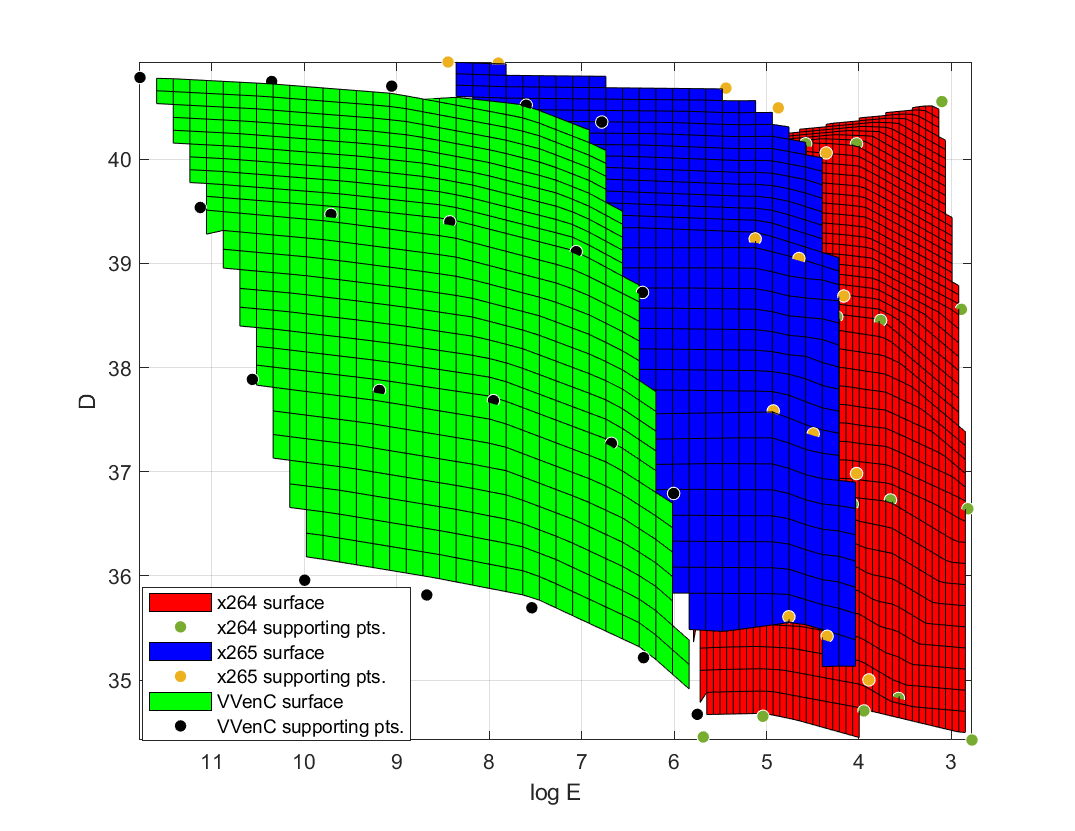}\par 
\end{multicols}
\vspace{-0.5cm}
\caption{(a) Rate-energy-distortion surfaces fitted using linear interpolation for the sequence "BasketballDrive," (b) Projection of the rate-energy-distortion surface to R-E plane, viewed from high PSNR values, (c) Projection of the rate-energy-distortion surface to E-D plane, viewed from high rate values, for all the encoders considered. In all these figures, the x264 surface is depicted using red, and supporting points are in green; the x265 surface is depicted using blue, and supporting points are in yellow-brown; and the VVenC surface is depicted using green, and supporting points are in black. In addition, D represents the distortion in terms of PSNR in dB, E represents the encoding energy in Joules, and R represents the rate in (kb/s). }
\vspace{-0.5cm}
\label{fig:3D}
\end{figure*}

\section{Evaluation}\label{sec:eval}
For evaluation, we employ the mean absolute percentage error (MAPE), a performance metric that gauges the disparity between actual and predicted data points in percentage terms. It is computed as follows:
\begin{equation} \label{eval}
\small
\overline{\epsilon_t} = 100 \cdot \frac{1}{N_t} \sum_{i=1}^{N_t} \left\vert \frac{\hat{d}_{t,i} - d_{t,i}}{d_{t,i}} \right\vert,
\end{equation}
where $N_t$ denotes the total number of data points. In this context, $t$ represents the categorization of data points into supporting (s) or non-supporting (ns), $\hat{d}_{t,i}$ signifies the distortion predicted through the surface fitting techniques discussed in Section \ref{sec:fitting}, and $d_{t,i}$ represents the actual distortion observed during encoding. Firstly, we assess the performance of the surface fitting techniques by evaluating their ability to accurately fit supporting (s) points. To do this, we use the $N_{s}$ supporting points, rate-energy-distortion triplets ($r_{\mathrm {s}}, e_{\mathrm {s}}, d_{\mathrm {s}}$) as mentioned in Section \ref{sec:setup}, where $N_{s}=20$ for x264, x265, and VVenC encoders.  

Secondly, similar to \cite{BB},  we analyze the performance of these techniques regarding the fitting accuracy for non-supporting (ns) data points. To achieve this, for x264 and x265 encoders, we generate a set of $N_{ns}= 24$ data points per sequence, for six intermediate CRF values $\{20,21,25,26,30,31\}$ and four intermediate presets $\textit{superfast}$, $\textit{faster}$, $\textit{medium}$, and $\textit{slower}$ forming the (intermediate) non-supporting points, defined by rate-energy-distortion triplets ($r_{\mathrm {ns}}, e_{\mathrm {ns}}, d_{\mathrm {ns}}$) on the surface. Subsequently, we use the generated surface to derive the fitted distortion values $\hat d_{\mathrm {ns}}$ for these non-supporting points based on their rate-energy data ($r_{\mathrm {ns}}, e_{\mathrm {ns}}$).

Table \ref{tab:accuracy} presents the MAPE values for surface fitting techniques utilized in the study, both for supporting and non-supporting points, averaged across all sequences. The results indicate that linear interpolation demonstrates the highest accuracy, with MAPE values of 0\% for supporting points across all encoders and 0.56\% and 0.48\% for non-supporting points in x264 and x265 encoder implementations, respectively. Following linear interpolation, the polynomial expression referenced by \eqref{poly32} shows relatively lower accuracy, recording MAPE values of 0.62\%, 0.40\%, and 0.11\% for supporting points in x264, x265, and VVenC encoder implementations, and 0.94\% and 0.64\% for non-supporting points in x264 and x265 encoder implementations, respectively. Lastly, the polynomial expression denoted by equation \eqref{custom} exhibits the lowest accuracy among all considered surface fitting techniques, with MAPE values of 1.03\%, 0.64\%, and 0.19\% for supporting points in x264, x265, and VVenC encoder implementations, and 1.15\% and 0.72\% for non-supporting points in x264 and x265 encoder implementations, respectively.

\section{Discussion}\label{sec:discussion}
Figure \ref{fig:3D}(a) illustrates three Rate-energy-distortion (R-E-D) surfaces for the sequence "BasketballDrive," representing distinct video encoders: x264 (depicted by a red surface with green supporting points), x265 (depicted by a blue surface with yellow-brown supporting points), and VVenC (depicted by a green surface with black supporting points). These surfaces provide a visual representation of the trade-offs between rate, encoding energy, and distortion for three encoders. As we traverse from left to right on each surface, presets transition from slow to fast, while moving from top to bottom signifies an increase in CRF/QP. By examining projections of these three 3-D surfaces, we can gain insights into the performance of the encoders across different rate, energy, and distortion scales. For example, the overlapping regions in the projections of these surfaces offer valuable depictions of the trade-offs among the different encoders.

Figure \ref{fig:3D}(b) depicts the R-E projections obtained from the surfaces that are illustrated in Figure \ref{fig:3D}(a). This projection illustrates the spectrum of rate-energy pairs achievable by different CRF/QP and preset configurations for all three encoders. From these R-E projections, the encoder with higher PSNR for the same rate-energy values is desired. Consequently, the overlapping regions indicate that the occluded encoder has a lower PSNR  for the same rate-energy performance and, thus, should not be chosen. The overlapping regions between x264 and x265 indicate that slow presets of the x264 encoder should not be used because x265 offers a higher quality at the same bitrate-energy performance. Similarly, the overlapping regions between x265 and VVenC indicate that slow presets of the x265 encoder should not be used because VVenC offers higher quality at the same bitrate-energy performance.

Figure \ref{fig:3D}(c) depicts the E-D projections obtained from the surfaces that are illustrated in Figure \ref{fig:3D}(a). This projection illustrates the spectrum of energy-distortion pairs achievable by different CRF/QP and preset configurations for all three encoders. From these R-E projections, the encoder with a lower rate for the same energy-distortion performance is desired. Accordingly, the overlapping regions indicate that the occluded encoder has a higher rate for the same energy-PSNR values and, thus, should not be used. The overlapping regions between x264 and x265 indicate that slow presets of the x264 encoder should not be used because x265 offers a lower rate at the same energy-PSNR level. Similarly, the overlapping regions between x265 and VVenC indicate that slow presets of the x265 encoder should not be used because VVenC offers a lower rate at the same energy-PSNR level. The behavior recorded in Figure \ref{fig:3D}(b) and Figure \ref{fig:3D}(c) is consistent throughout the dataset used in this study. 

\section{Conclusion}
This work addresses a significant gap in video coding evaluation methods concerning energy consumption by introducing a 3D representation that incorporates rate, encoding energy, and distortion using surface fitting techniques. Through the analysis of various surface fitting approaches, our results highlight the effectiveness of linear interpolation in accurately representing rate-energy-distortion surfaces. Additionally, we explore the trade-offs between rate, energy, and distortion across different encoder configurations and quality levels using the fitted surfaces. Lastly, we demonstrate the practical application of this 3D representation through examples of projected representations, which further shows that the slow presets of the older encoders (x264, x265) should be avoided, as the recent encoders (x265, VVenC) offer a higher quality for the same bitrate-energy performance and provide a lower rate for the same energy-distortion performance. In the future, we aim to investigate detailed studies on content dependencies of these projections, and extend the 3D representations to other distortion metrics such as SSIM and VMAF.
\section*{Acknowledgment}
This work was funded by the Deutsche Forschungsgemeinschaft (DFG, German Research Foundation) – Project-ID 447638564.
\bibliographystyle{IEEEbib}
    \bibliography{literature}
\color{red}
\end{document}